%% file: 2017PDF-physical.v2.tex
\documentclass[prd,aps,twocolumn,preprintnumbers, showpacs, nofootinbib,superscriptaddress,notitlepage]{revtex4-1}
\usepackage{amssymb,amsmath,graphicx,hyperref}
\usepackage{mathrsfs}
\usepackage{amsfonts}
\usepackage{amsmath}
\usepackage{slashed}
\usepackage{array}
\usepackage{verbatim}
\usepackage{epsfig}
\usepackage{graphicx}
\usepackage{color}
\usepackage[dvipsnames]{xcolor}

\newcommand{\beq}{\begin{eqnarray}}
\newcommand{\eeq}{\end{eqnarray}}

\newcommand{\nn}{\nonumber}

\DeclareMathOperator\erf{erf}

\begin{document}

\title{Improved Parton Distribution Functions at Physical Pion Mass}

\author{Huey-Wen Lin}
\email{hwlin@pa.msu.edu}
\affiliation{Department of Physics and Astronomy, Michigan State University, East Lansing, MI 48824}
\affiliation{Department of Computational Mathematics,
  Science and Engineering, Michigan State University, East Lansing, MI 48824}

\author{Jiunn-Wei Chen}
\affiliation{Department of Physics, Center for Theoretical Sciences, and Leung Center for Cosmology and Particle Astrophysics, National Taiwan University, Taipei, Taiwan 106}
\affiliation{Helmholtz-Institut f\"{u}r Strahlen- und Kernphysik and Bethe Center
for Theoretical Physics, Universit\"{a}t Bonn, D-53115 Bonn, Germany}

\author{Tomomi Ishikawa}
\affiliation{T.~D.~Lee Institute, Shanghai Jiao Tong University, Shanghai, 200240, P. R. China}

\author{Jian-Hui Zhang}
\affiliation{Institut f\"{u}r Theoretische Physik, Universit\"{a}t Regensburg, D-93040 Regensburg, Germany}

\collaboration{LP$^3$ Collaboration}
\preprint{MSUHEP-17-014}

\begin{abstract}
{
We present the first lattice results on isovector unpolarized and longitudinally polarized parton distribution functions (PDFs) at physical pion mass. The PDFs are obtained using the large-momentum effective field theory (LaMET) framework where the full Bjorken-$x$ dependence of finite-momentum PDFs, called ``quasi-PDFs'', can be calculated on the lattice. The quasi-PDF nucleon matrix elements are renormalized nonperturbatively in RI/MOM-scheme. 
However, the recent renormalized quasi-PDFs suffer from unphysical oscillations that alter the shape of the true distribution as a function of Bjorken-$x$. In this paper, we propose two possible solutions to overcome this problem, and demonstrate the efficacy of the methods on the 2+1+1-flavor lattice data at physical pion mass with lattice spacing 0.09~fm and volume $(5.76\mbox{ fm})^3$. 
}

\end{abstract}

\maketitle

Parton distribution functions (PDFs) are universal nonperturbative properties of the nucleon which describe the
probability densities of quarks and gluons seen by an observer moving at the speed of light relative to the hadron. 
The unpolarized PDFs can be used as inputs to predict cross sections in high-energy scattering experiments, one of the major successes of QCD. 
These distributions can be extracted through global analysis using multiple experiments by factorizing 
the hard-scattering cross sections into the PDFs and short-distance matrix elements that are calculable in perturbation theory. 
Today, multiple collaborations provide regular updates concerning the phenomenological determination of the PDFs~\cite{Ball:2012cx,Ball:2014uwa,Harland-Lang:2014zoa,Dulat:2015mca,Alekhin:2017kpj,Owens:2012bv} using the latest experimental results, from medium-energy QCD experiments at Jefferson Lab in the US to high-energy collisions at the LHC in Europe. 
Progress has also been made in polarized PDF using data from RHIC at Brookhaven National Lab, the COMPASS experiment at CERN; further progress would be made at the proposed electron-ion collider (EIC). 
After the past half century of effort in both theory and experiment, our knowledge of PDFs has greatly advanced. However, as the experiments get more precise, the precision needed in PDFs to make Standard-Model (SM) predictions has increased significantly too. Current and planned experiments (such the EIC) will go further into unexplored or less-known regions, such as sea-quark and gluonic structure. 
We would like to explore these unknown regions using first-principles calculation from the Standard Model in lattice QCD.

A new approach for the direct computation of the $x$-dependence of PDFs on the lattice was proposed by Ji~\cite{Ji:2013dva,Ji:2014gla}: large-momentum effective theory (LaMET), where lightcone PDFs can be obtained by approaching the infinite-momentum frame (IMF). 
Prior to this development, lattice QCD was limited to calculating charges and various leading moments of nucleon matrix elements, integrals of the PDFs, through the operator product expansion (OPE).
In LaMET on the lattice, we start by calculating the ``quasi-PDF'' $\tilde{q}$ with a nucleon moving along the $z$ direction with finite-momentum $P_z$. The quasi-PDF is an integral over matrix elements $h$ with a spatial correlation of partons,
\begin{equation}
\tilde{q}(x, P_z, \tilde{\mu}) 
= \int_{-\infty}^\infty \frac{dz}{2\pi}\ e^{ixP_zz} h(z,P_z,\tilde{\mu}) \,,
\label{eq:quasipdf}
\end{equation} 
where  $\tilde{\mu}$ is the renormalization scale in a chosen renormalization scheme. 
The boosted-nucleon matrix elements are
\begin{equation} 
h(z,P_z,\tilde{\mu}) 
= \frac{1}{2} \big\langle P \big| \bar{\psi}(z) \Gamma \exp\left(ig\int_0^z dz' A_z(z') \right)  \psi(0) \big|P\big\rangle
\,, 
\label{eq:NME}
\end{equation} 
for $\Gamma=\gamma_z$ or $\gamma_t$ for unpolarized case. 
To obtain polarized PDFs, $\Delta\tilde{q}$, we calculate $\Delta h(z)$ as in Eq.~\ref{eq:NME} with $\Gamma=\gamma_5\gamma_z$.
Note that the quasi-PDF depends nontrivially on the nucleon momentum $P_z$, unlike the lightcone PDF.  Within LaMET framework, the lightcone PDFs can be 
matched to the quasi-PDF through a factorization formula~\cite{Ji:2013dva,Ji:2014gla}:
\begin{align} \label{eq:factorization}
\tilde{q}(x, P_z, \tilde{\mu}) &= \int_{-1}^{+1} \frac{dy}{|y|} \ C\left(\frac{x}{y}, \frac{\tilde{\mu}}{P_z},\frac{\mu}{P_z}\right) q(y,\mu)
\nn\\ 
&\quad
+ {\cal O}\bigg(\frac{M_N^2}{P_z^2}, \frac{\Lambda_{\text{QCD}}^2}{P_z^2} \bigg)
\ ,
\end{align}
where $M_N$ is the nucleon of mass, $C$ is the matching kernel, and the ${\cal O}(M_N^2/ P_z^2, \Lambda_{\text{QCD}}^2/ P_z^2)$ terms are power corrections suppressed by the nucleon momentum. 
Here, $q(y,\mu)$ for negative $y$ corresponds to the antiquark contribution. 
Refs.~\cite{Ji:2013dva,Ji:2014gla,Ma:2014jla}
show $\tilde{q}$ and $q$ have the same infrared (IR) divergences; therefore, the matching kernel $C$ only depends on ultraviolet (UV) physics and can be calculated in perturbative QCD~\cite{Xiong:2013bka,Ma:2014jla}.
Since Ji's proposal in 2013, there have been many follow-up works concerning the quasi-PDFs. 
On the lattice side, there have been
many lattice-QCD calculations of the nucleon isovector quark distributions~\cite{Lin:2014zya,Alexandrou:2015rja,Chen:2016utp,Alexandrou:2016jqi,Monahan:2016bvm,Orginos:2017kos,Chen:2017mzz,Green:2017xeu} 
including the unpolarized, polarized and transversity cases and variations of the quasi-PDF methods.  
Recently, there have also been a number of works on quasi-PDFs renormalization~\cite{Chen:2017mzz,Ji:2017oey,Green:2017xeu,Ishikawa:2017faj,Alexandrou:2017huk,Constantinou:2017sej,Xiong:2017jtn,Chen:2016fxx,Ishikawa:2016znu}.

\begin{figure*}[htbp]
\includegraphics[width=.4\textwidth]{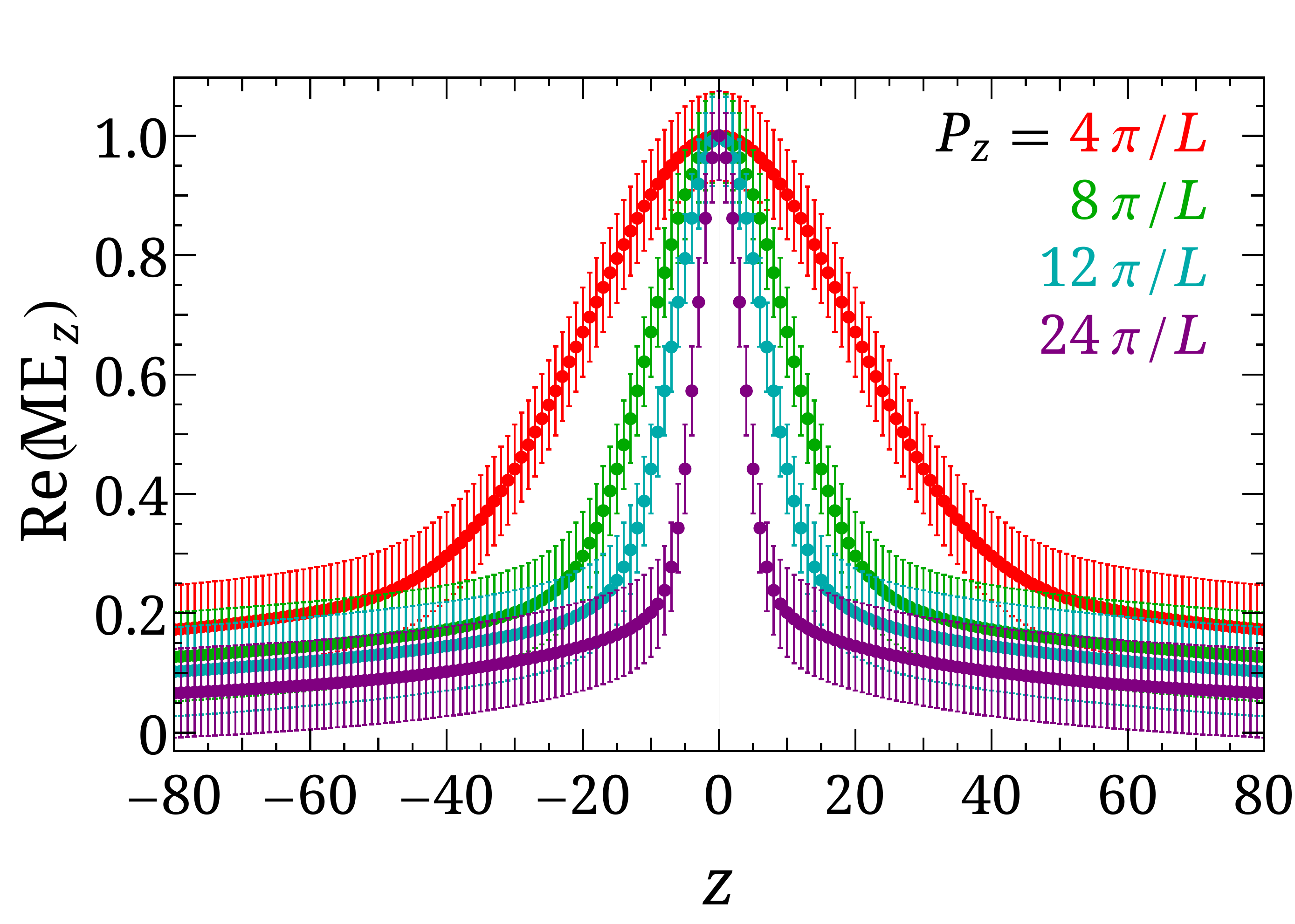}
\includegraphics[width=.4\textwidth]{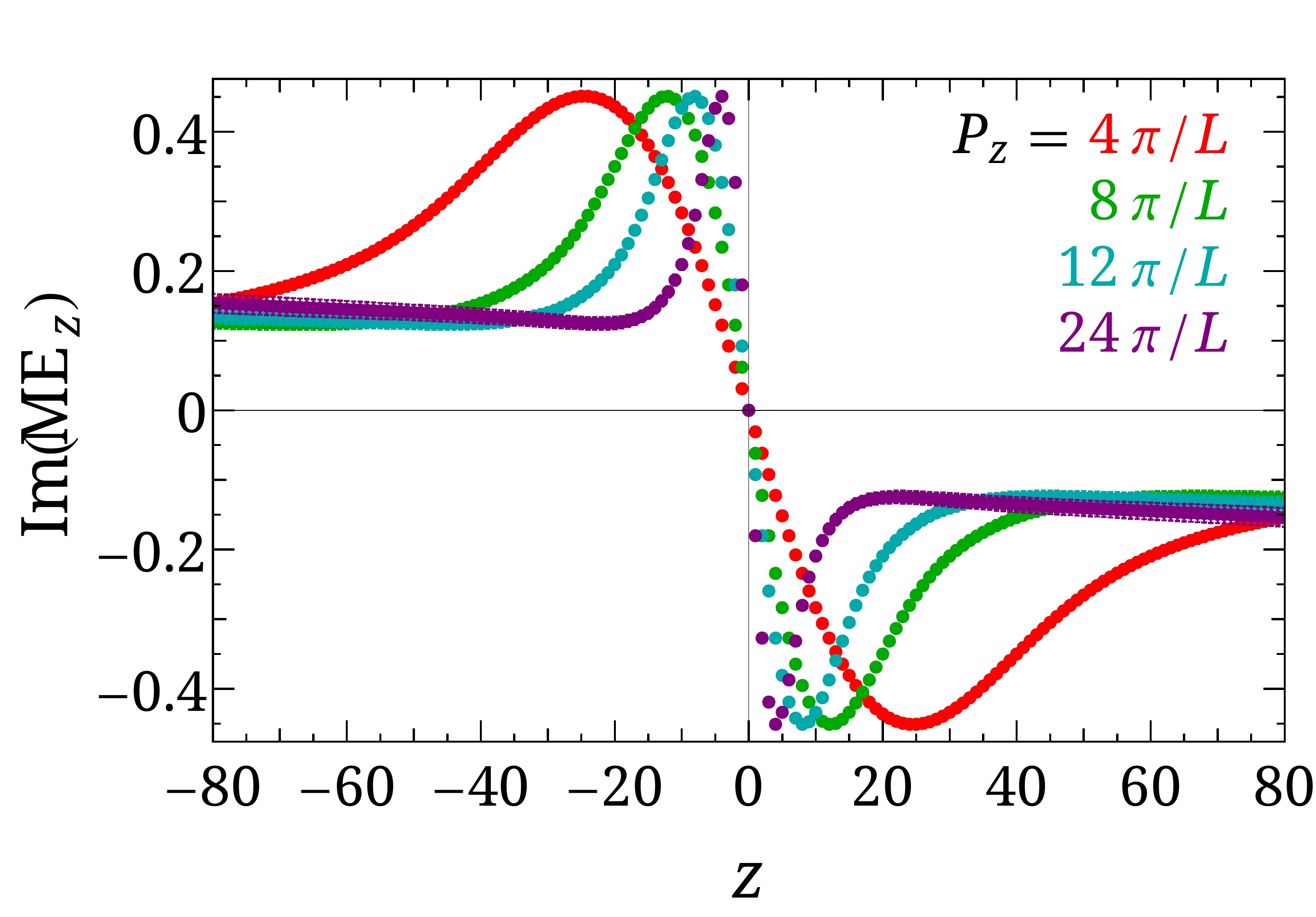}
\includegraphics[width=.4\textwidth]{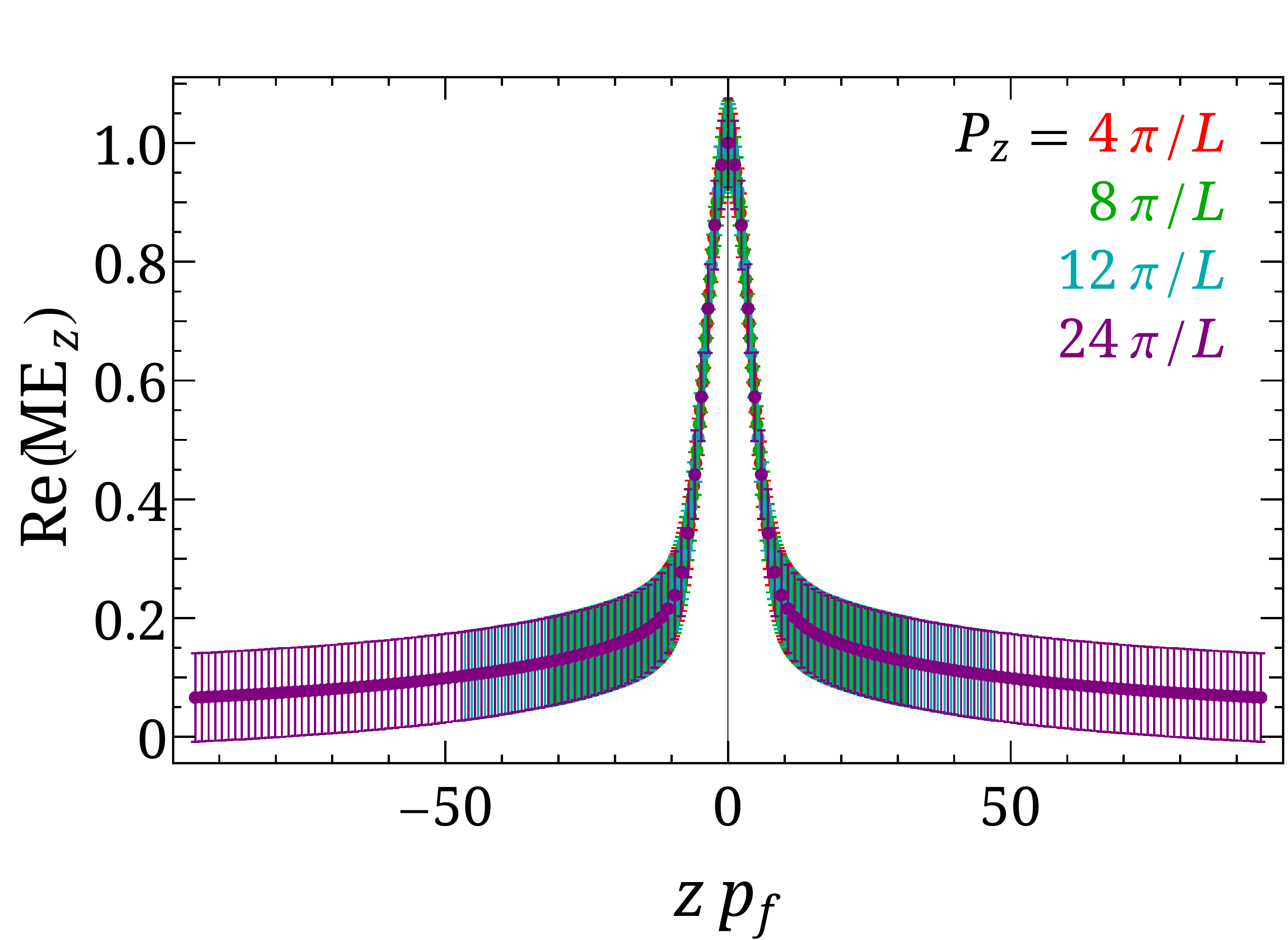}
\includegraphics[width=.4\textwidth]{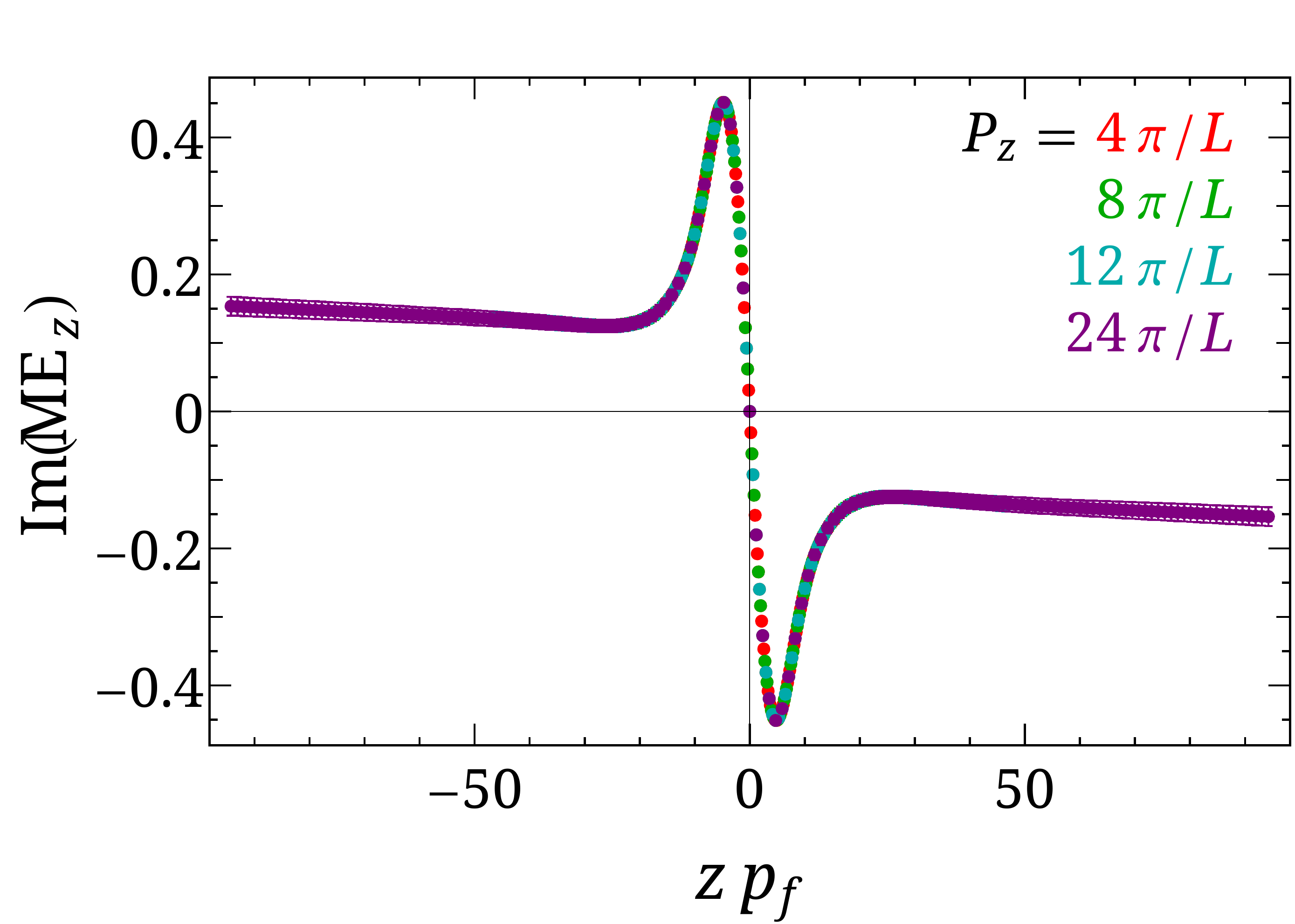}
\caption{The real (top left) and imaginary (top right) pseudo-nucleon matrix elements derived from CT14 PDFs as functions of $z$ with nucleon boosted momenta $P_z=\{4,8,12,24\}\pi/L$.
(Bottom) Similar plots as the top row but plotted along the x-axis in dimensionless units of $z P_z$. 
} 
\label{fig:hCT14}
\end{figure*}
Recent renormalized quasi-PDF studies have shown oscillatory 
behavior~\cite{Chen:2017mzz,Green:2017xeu} that is significantly 
distorting the true distribution, especially in the antiquark region. 
This happens due to the renormalization constants' enhancement of the large-$z$ matrix elements 
through exponential counter-term; in the bare quasi-PDF it is less severe. 
This issue was briefly mentioned in Ref.~\cite{Chen:2017mzz}
and the $z$-truncated and inverse Fourier-transformed central value of the CJ15 PDFs~\cite{Accardi:2016qay},
nucleon matrix elements $h(z)^\text{CJ15}$ are compared with the RI/MOM renormalized 
lattice ones $h(z)^\text{lattice}$. 
Due to the nature of the the global-fit parametrization of $x f(x)$, where $f$ is the valence up/down distribution
$\{u,d\}^v(x)$ with $f(x)$ parametrized as $x^{a_n} (1-x)^{b_n} P(x)$ 
and $P(x)$ a polynomial in $x$, when one takes the isovector combination, 
$x(u(x)-d(x))$ (or the antiquark combination), it does not necessarily go to 
zero as individual quark components would. As a result, it is not surprising to see 
divergence in $u(x)-d(x)$ for $x$ near 0, and the same is true for the
antiquark distribution. Due to this behavior, $h(z)^\text{CJ15}$ has nonvanishing values 
at large $z$ far beyond the size of the largest currently available lattices. 
Similarly, the matrix elements $h(z)^\text{CJ15}$ near $z=0$ correspond to the
large-$x$ region, where most global analyses either rely on extrapolation or have to 
handle nuclear corrections, as CJ15 does; this introduces theory uncertainty to the global analysis. 
It was suggested in Ref.~\cite{Chen:2017mzz} that the lattice data may have
mixing with higher-twist operators in the large-$z$ region and that going to higher nucleon 
boosted momenta $P_z$ would reduce the $z$ range needed to reconstruct the quasi-PDFs.

In this paper, we take the CT14 NNLO PDF~\cite{Dulat:2015mca} where the errors for the isovector quark distribution have been properly taken into account and investigate these issues.
First, we transform the CT14 PDF without any mass correction to $h(z)^\text{CT14}$ 
using momenta $P_z=\{4,8,12,24\}\pi/L$ with $L=5.76$~fm, the lattice spatial length. 
Figure~\ref{fig:hCT14} shows the error in the PDF propagated accordingly, and we see the expected nonvanishing 
matrix elements at large $z$, especially the imaginary matrix elements, as expected. 
In the bottom rows of Fig.~\ref{fig:hCT14}, we see similar matrix elements but plotted as a function of 
$z P_z$ (as in Ref.~\cite{Chen:2017mzz}); 
since there is no nucleon-mass correction nor higher-twist effects here, the data points all 
lie on top of each other. 
If the small-$x$ region of the PDF is really as divergent as the global PDF extrapolation, 
we will need an alternative approach to be able to calculate small-$x$ PDFs in lattice calculations.

\begin{figure*}[htbp]
\includegraphics[width=.4\textwidth]{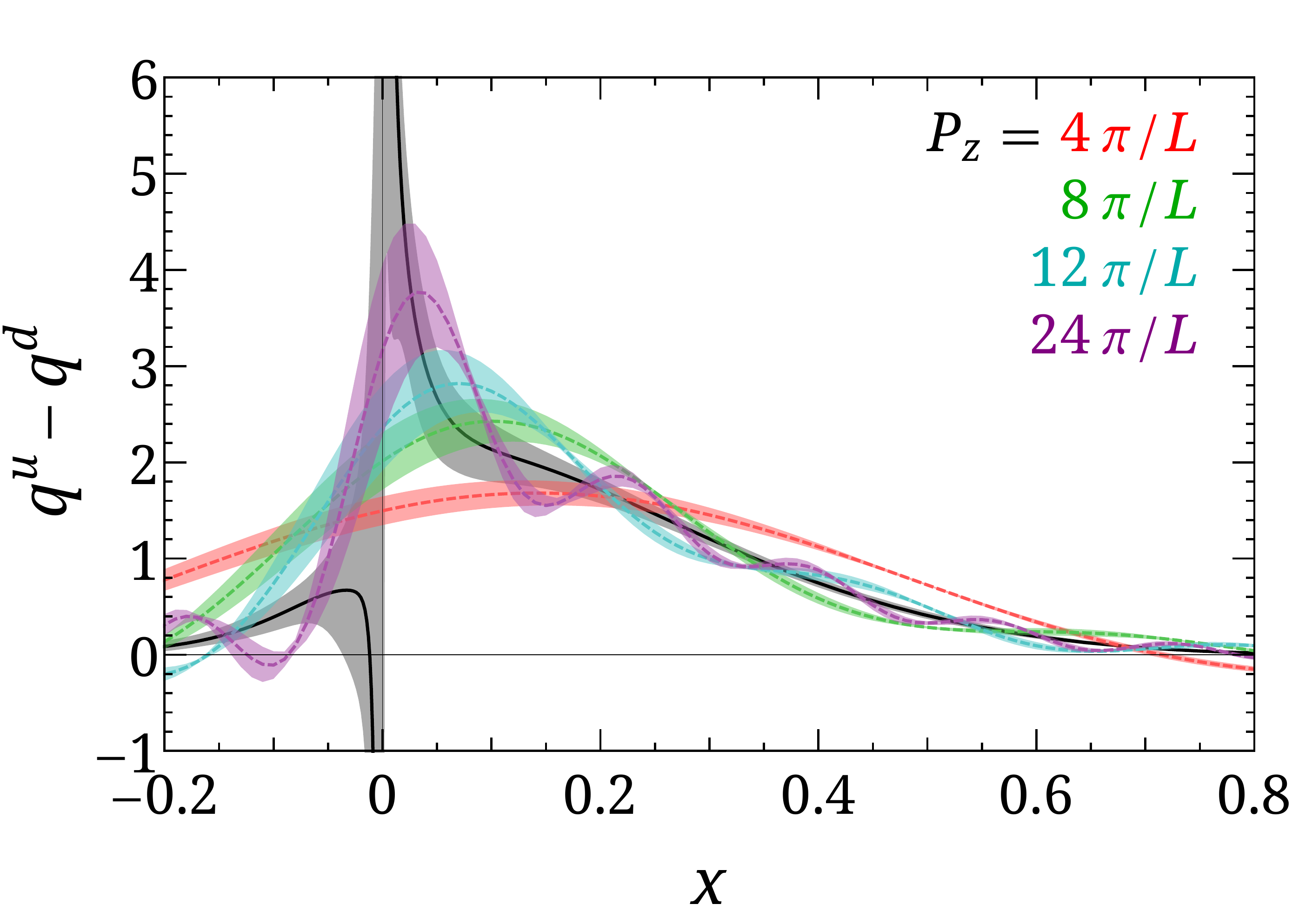}
\includegraphics[width=.4\textwidth]{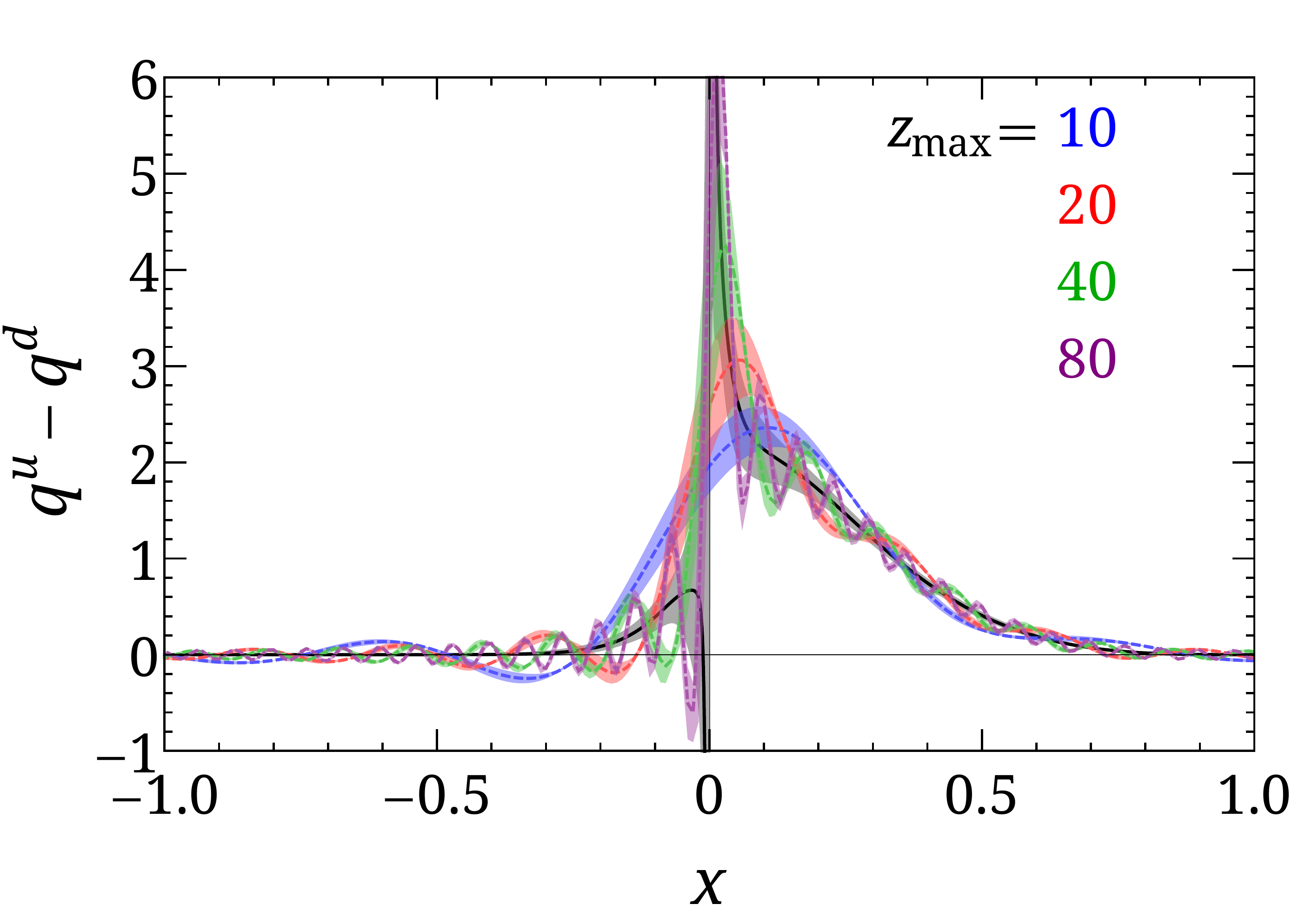}
\caption{(Left) The quasi-PDFs obtained from the matrix elements shown in Fig.~\ref{fig:hCT14} with $z_\text{max}=32$. The unphysical oscillations worsen as the nucleon momentum $P_z$ increases. This presents a serious problem: we need larger $P_z$ to recover the lightcone limit, but the unphysical oscillations alters the PDF shape.
(Right)  The quasi-PDFs obtained from the matrix elements with $P_z=24\pi/L$ with different values of $z_\text{max}$. The larger values of $z_\text{max}$ have worse oscillations but possess the information needed to reach the smaller-$x$ region. 
 } 
\label{fig:qPDFCT14}
\end{figure*}

Using the ideal pseudo-data $h(z)^\text{CT14}$, we can start to investigate the sources of the 
unphysical oscillations in the lattice quasi-PDFs and seek ways to improve or remove them. 
First, we study the oscillatory behavior as functions of the boosted momentum inputs.  
The left-hand side of Fig.~\ref{fig:qPDFCT14} shows the transform of $h(z,P_z)^\text{CT14}$ using the quasi-PDF formulation 
with a $z$ cutoff of 32, half of the lattice spatial size we used at the physical pion mass. 
One can clearly see that the oscillatory behavior worsens as the $P_z$ used in the calculation increases. 
This also indicates that the oscillations are purely an artifact coming from the truncation of the Fourier transformation.
Let us focus on the oscillatory behavior of the PDFs at the largest $P_z=24\pi/L$ here. 
We transform the $h(z,P_z=24\pi/L)^\text{CT14}$ using the quasi-PDF formulation
with different $z_\text{max}$ cutoffs $\{10,20,40,80\}$, as shown in the right-hand side of Fig.~\ref{fig:qPDFCT14}. 
The smaller-$z$ cutoffs have milder oscillatory contamination of the resulting PDFs; however, the larger-$z$ cutoff recovers
better small-$x$ PDFs though the oscillation is more severe. We need a better way to
recover as much of the true distribution as possible. 

\begin{figure*}[htbp]
\includegraphics[width=.4\textwidth]{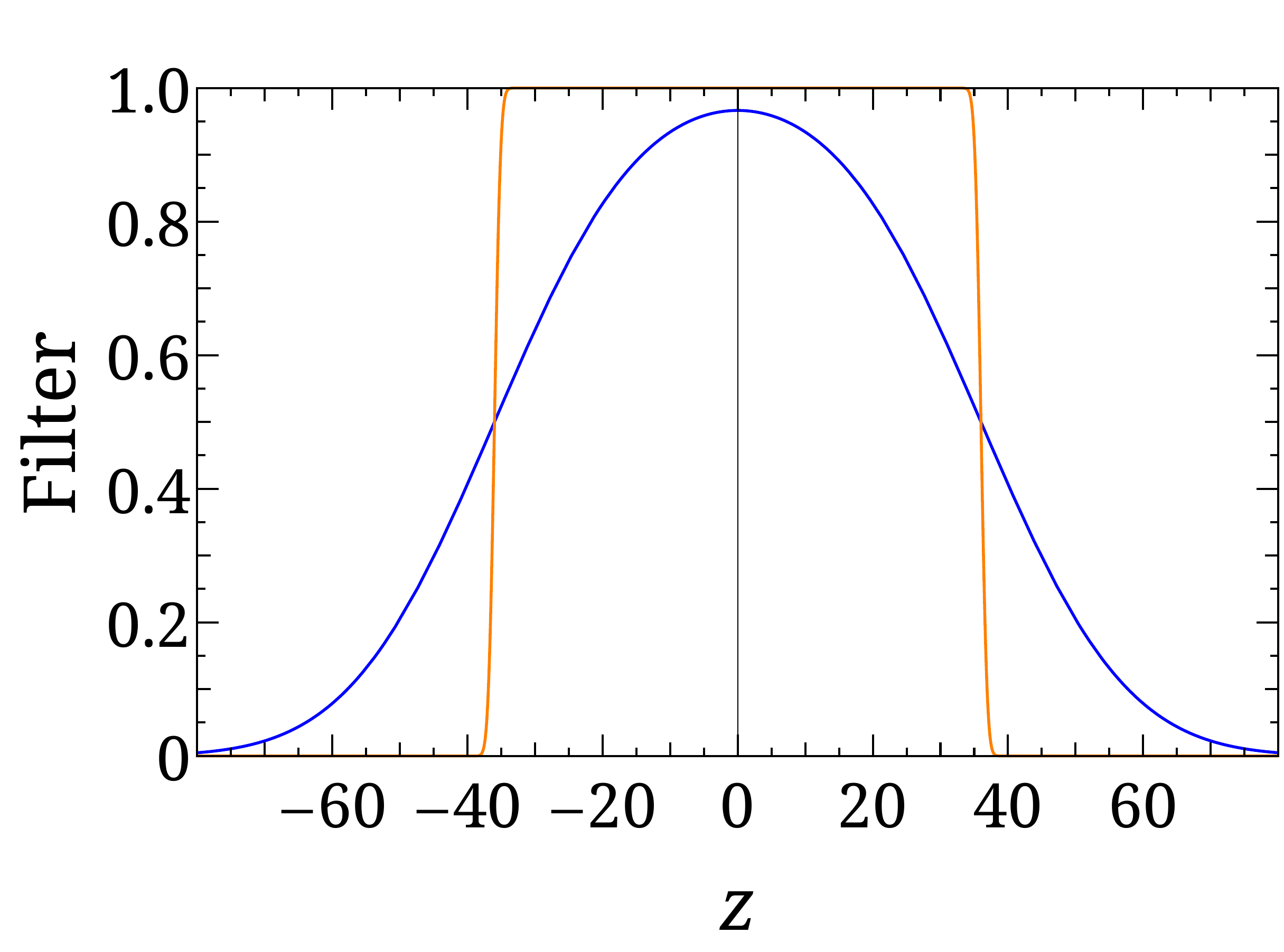}
\includegraphics[width=.4\textwidth]{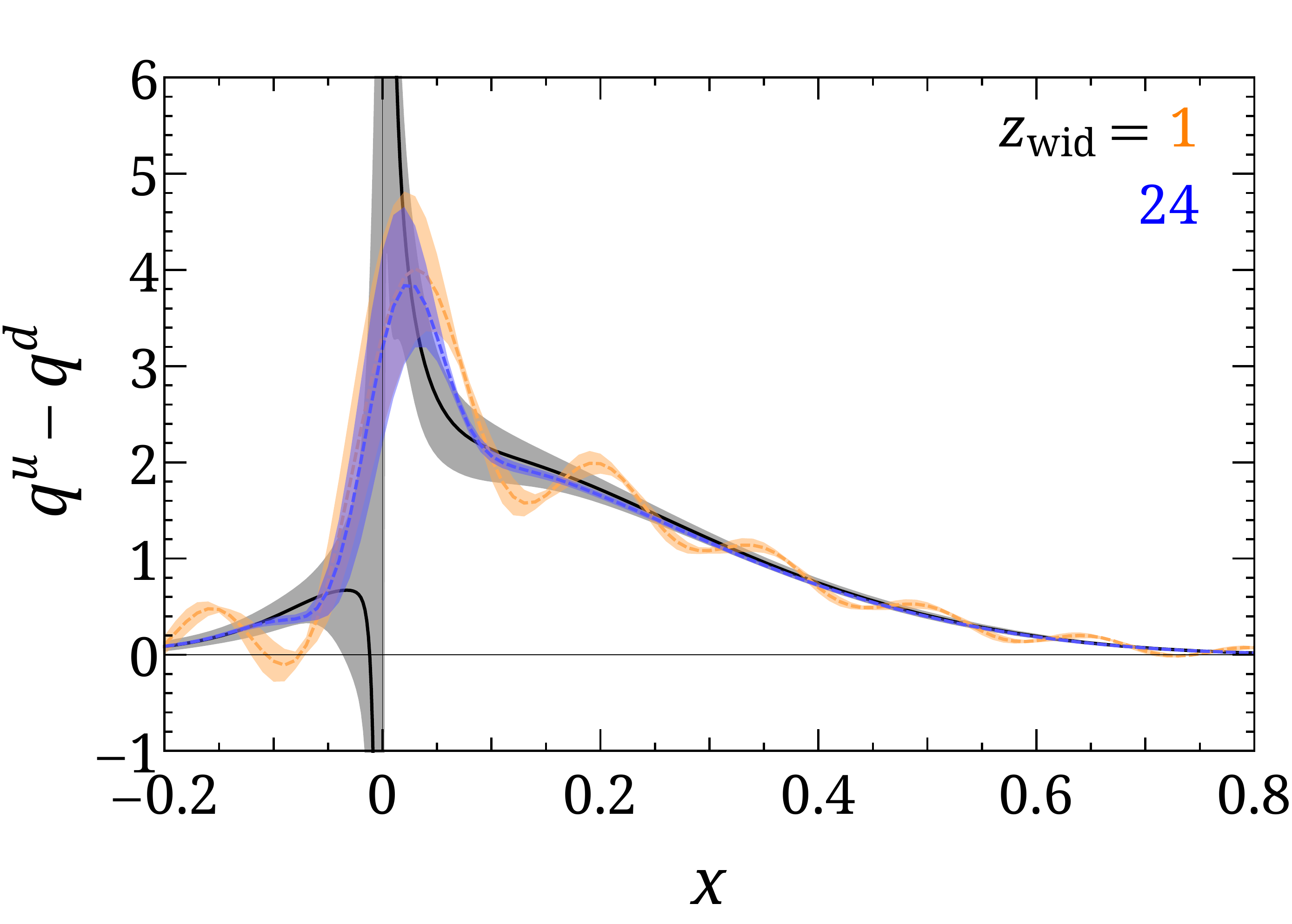}
\includegraphics[width=.4\textwidth]{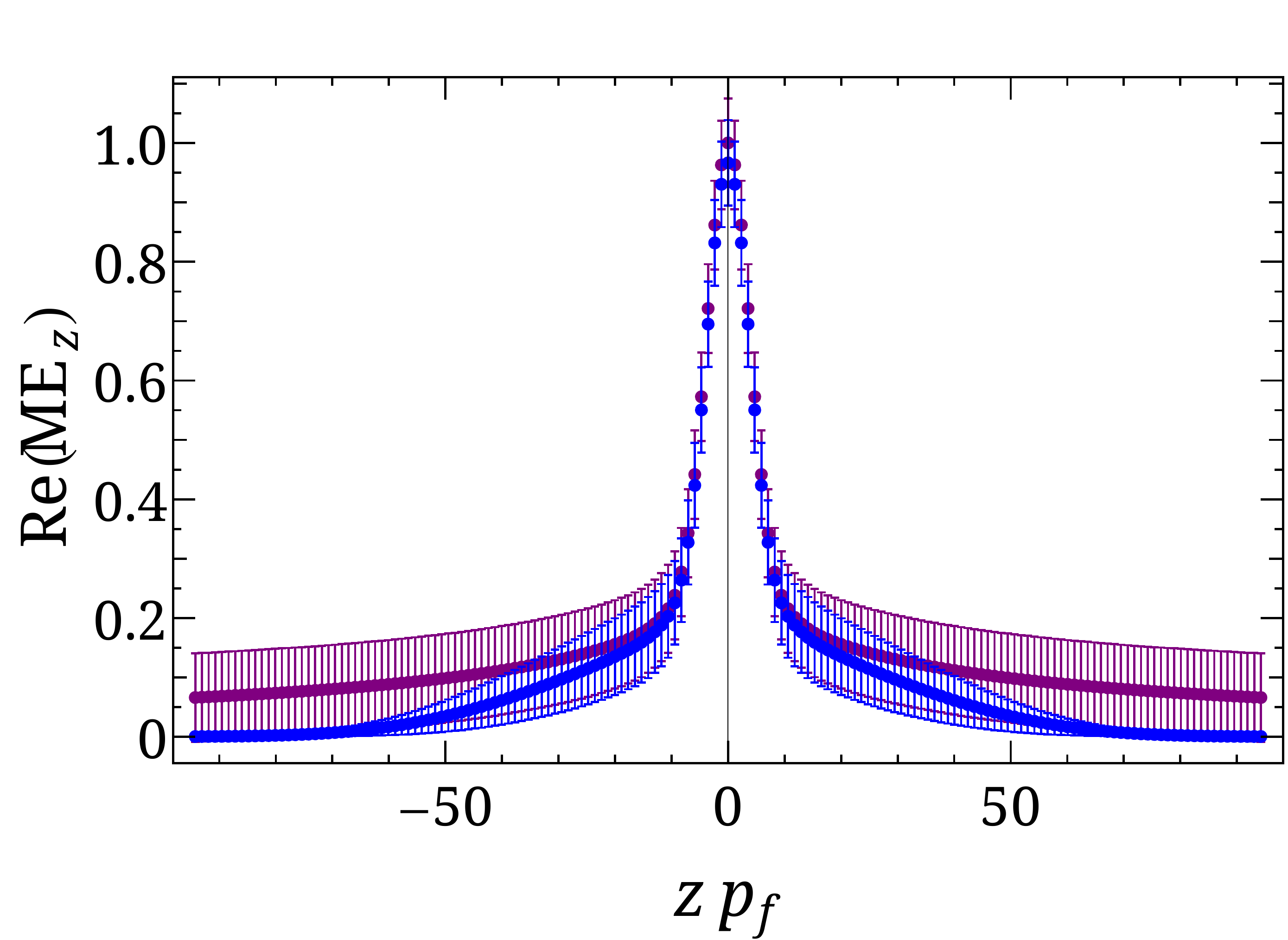}
\includegraphics[width=.4\textwidth]{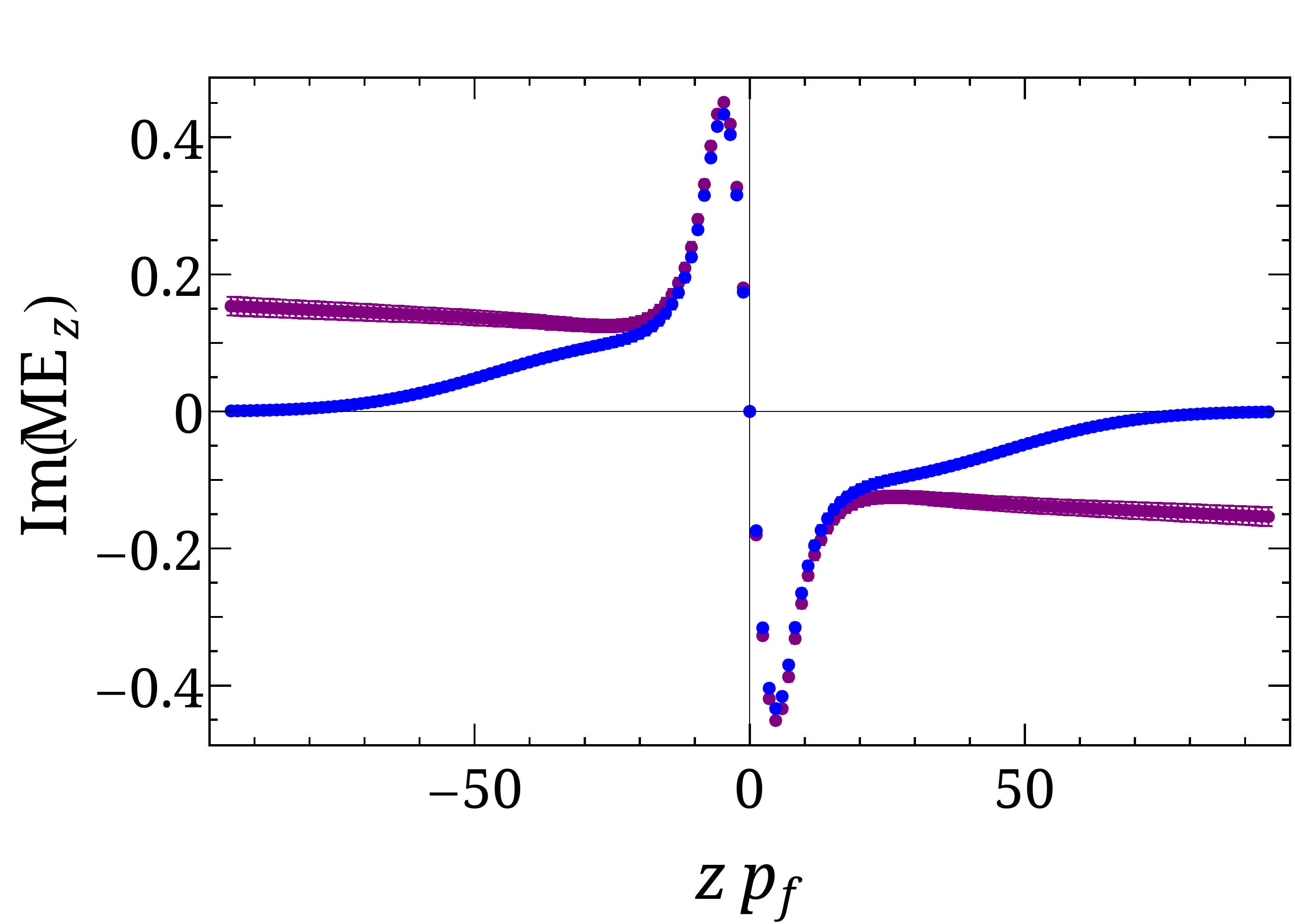}
\caption{(Top left) The filter shapes proposed to apply to $h(z)$ with $\{z_\text{max},z_\text{wid}\}=\{36,24\}$ (blue) and $\{32,1\}$ (orange); the latter is the same as a hard $z$ cutoff. The real (bottom left) and imaginary (bottom right) matrix elements before (purple) and after (blue) applying the $F(36,24)$ filter. (Top right) The soft-filtered PDF (blue) has significant improvement in recovering the original PDF (gray) while the hard cutoff (orange) suffers from significant unphysical oscillations throughout the entire $x$ range. } 
\label{fig:filters}
\end{figure*}

One solution proposed here is to remove these oscillations by means of a low-pass filter. Note that 
such a filter should be $P_z$ dependent, and one can carefully tune the 
function that best recovers the true PDF from a selected set of PDFs as guidelines. 
Here, we demonstrate one possible filter formulation; one should keep in mind that there are many possible 
solutions, and we should continue exploring to find the most robust way to remove the unphysical oscillations from the 
quasi-PDFs. 
We propose a hat-shaped filter in $z$ constructed using the sigmoidal error functions:
\begin{align}
\label{eq:filter}
F(z_\text{lim}, z_\text{wid}) = \frac{1+\erf\left(\frac{z+z_\text{lim}}{z_\text{wid}}\right)}{2} \frac{1-\erf\left(\frac{z-z_\text{lim}}{z_\text{wid}}\right)}{2} 
\end{align}
For $F(36,24)$, the filter shape is shown in the upper-left corner of Fig.~\ref{fig:filters}. For comparison, we show in orange the filter $F(32,1)$, which is close to a simple $z_\text{max}$ cutoff in the quasi-PDFs transformation. 
The lower rows of Fig.~\ref{fig:filters} show how the filter alters the matrix elements; the smoother transition to zero at larger $z$ plays an important role in removing the unphysical oscillation.
The quasi-PDFs corresponding to these filtered matrix elements are shown in the upper-right of Fig.~\ref{fig:filters}.
There is definite a reduction in the oscillatory behavior and 
significant improvement in recovering the medium to large-$x$ region of the PDFs. One can recover the PDFs up to $x=0.1$ region. 

We expect to test the filter function on the known PDFs through similar exercises in this work, then use the same function on lattice data; the mismatch region between the original PDFs and the quasi-PDFs tells us where lattice systematics are present.

\begin{figure*}[htbp]
\includegraphics[width=.4\textwidth]{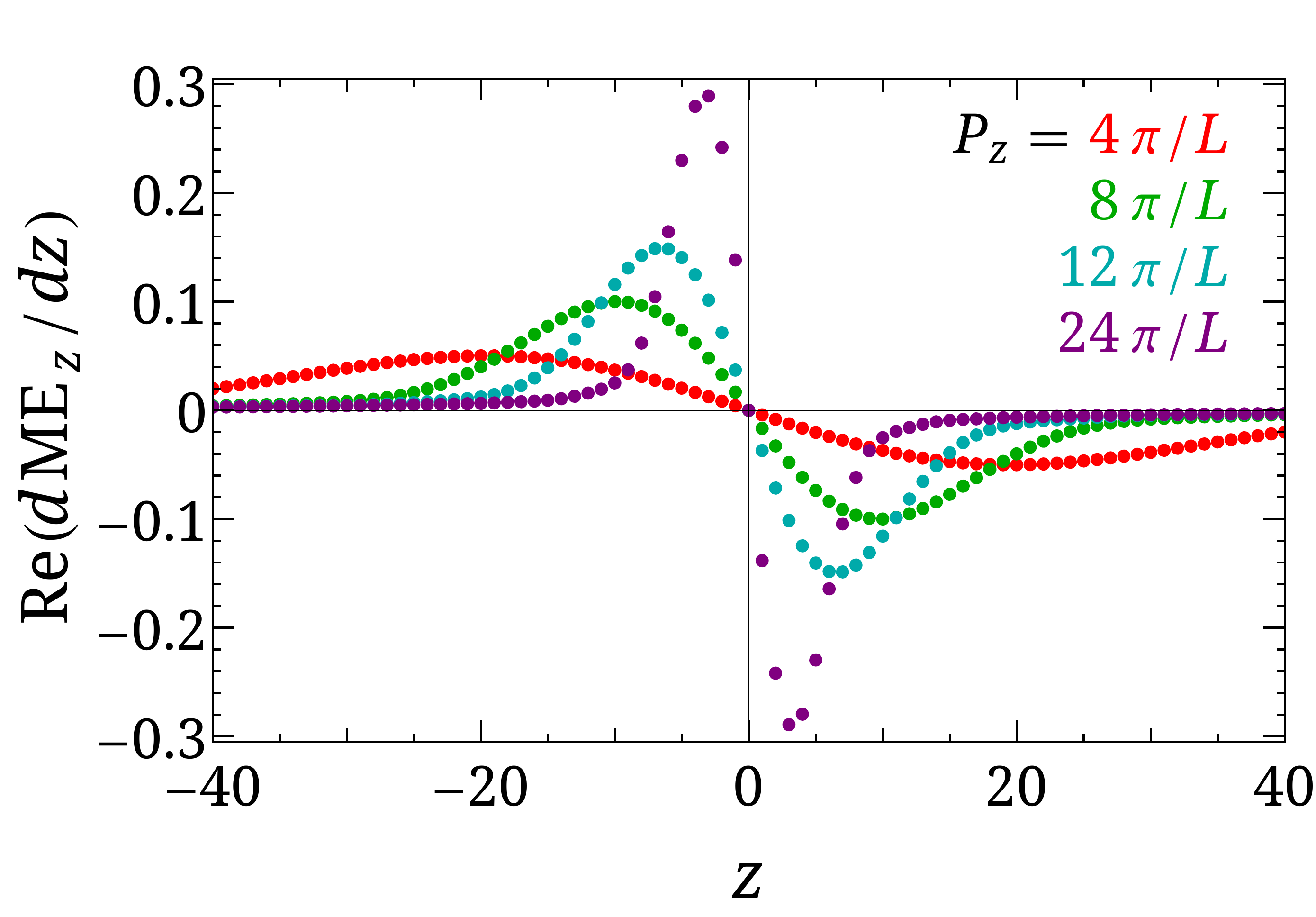}
\includegraphics[width=.4\textwidth]{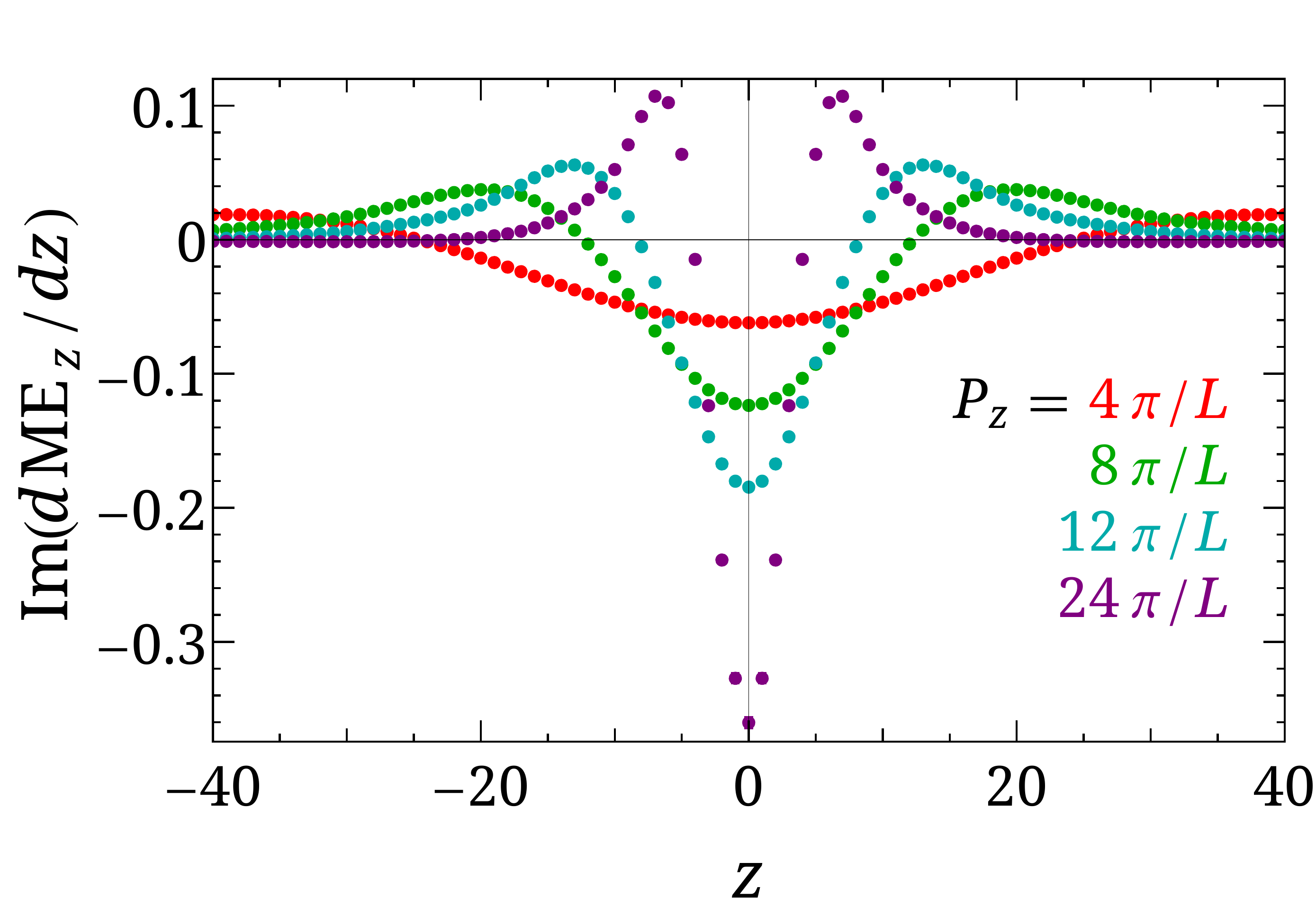}
\includegraphics[width=.4\textwidth]{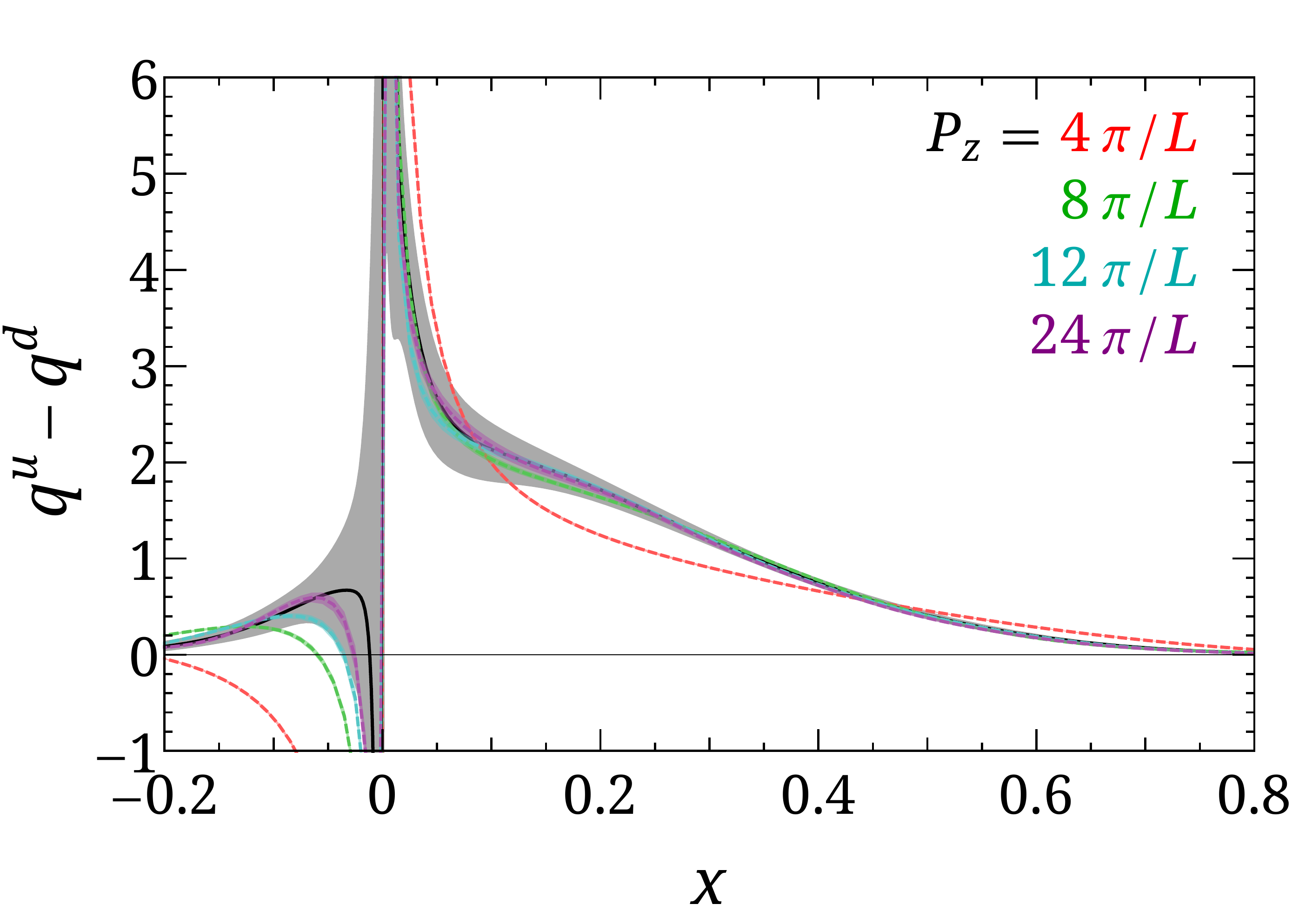}
\includegraphics[width=.4\textwidth]{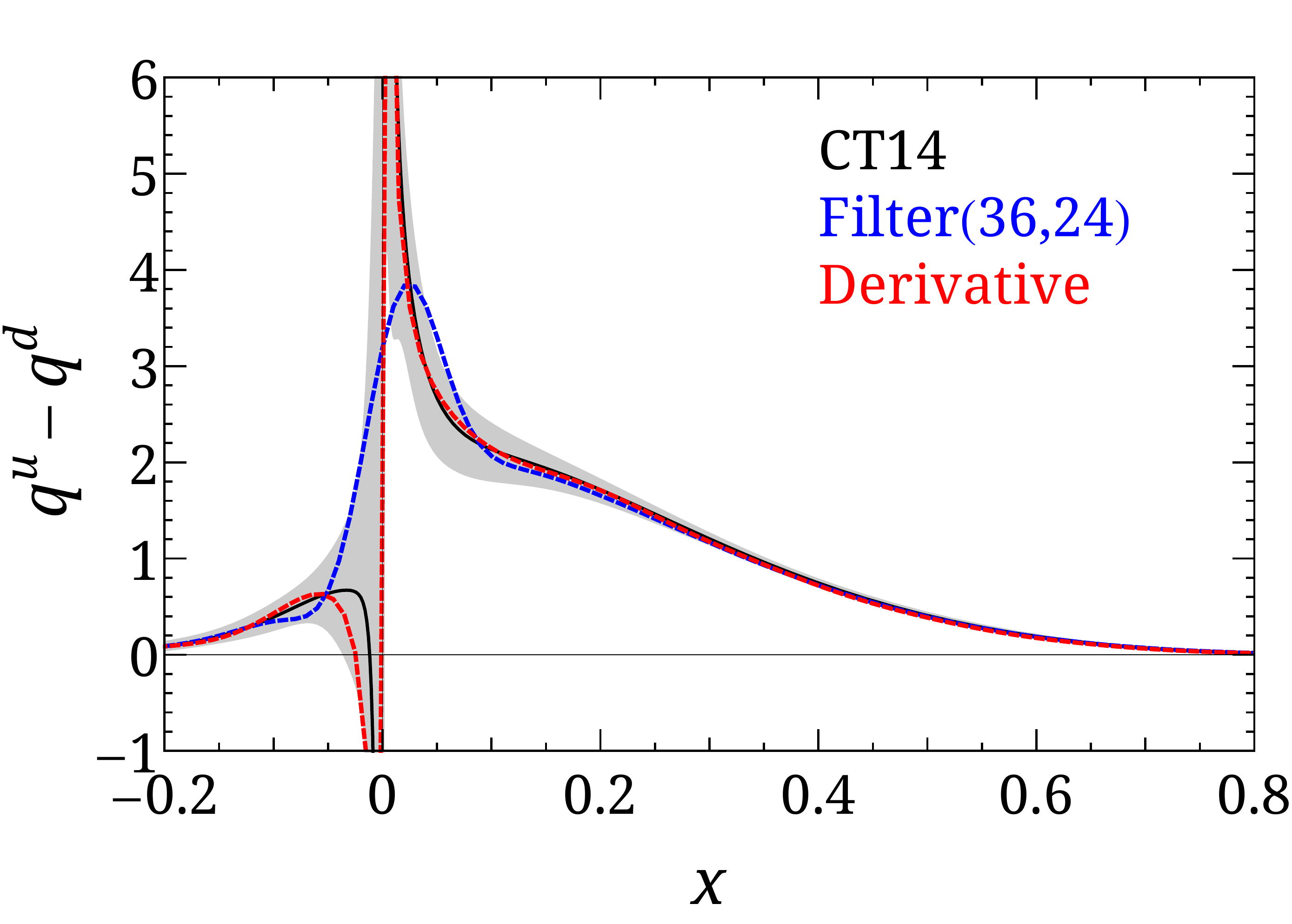}
\caption{The real (top left) and imaginary (top right) components of $h^\prime(z)$ as a function of $z$ with $P_z=\{4,8,12,24\}\pi/L$. 
(Bottom left) The original CT14 PDF (gray) along with the new PDFs obtained from the ``derivative'' method with $P_z=\{4,8,12,24\}\pi/L$ using $F(32,1)$. The oscillatory behavior is gone and positive $x$ is mostly recovered for $P_z > 4\pi/L$. However, for the antiquark (negative $x$) region, with $P_z =24\pi/L$ one can trust up to $|x| \approx 0.05$. 
(Bottom tight) The original CT14 PDF (gray) in comparison with the recovered PDF with $P_z=24\pi/L$ using the filter method described earlier (blue) and derivative method (red). Both methods recover the distribution in the $|x| > 0.1$ region; only the derivative method can recover the divergence in the small-$x$ region with $z_\text{max}=32$.} 
\label{fig:derivative}
\end{figure*}

Another proposal suggested in this paper is the ``derivative'' method; this should not only remove the unphysical 
oscillation but also allow us to 
reach smaller $x$ for the PDFs. This is essential, since the current knowledge of the 
$x<0.001$ region may not be reliable; if there is indeed a divergence in the true distribution,
the lattice data should be able to address it without compromising the predictive power. 
For example, most of our nucleon matrix elements $h(x)^\text{lattice}$ go to zero at large $z$; 
if we manually set the large-$z$ value to $h(z,P_z)^\text{CT14}$ or parametrize 
in similar manner, we will only reproduce the PDF results. 
To take into account the potential nonvanishing of $h(z)$ at infinite values of $z$, we take the 
derivative of the nucleon matrix elements $h^\prime(z) = (h(z+1)-h(z-1))/2$. The Fourier expansion of this derivative differs from the original in a known way:
\begin{align}
\label{eq:derivative}
q(x) = \int_{-z_\text{max}}^{+z_\text{max}} \!\!\! dz \frac{-1}{2\pi} \frac{e^{i x P_z z}}{i P_z x} h^\prime(z)
\end{align}
The upper row of Fig.~\ref{fig:derivative} shows real (left) and imaginary (right) parts of $h^\prime(z)$ as functions of the nucleon momentum $P_z$. Here, we drop the error propagation of the PDFs, since the correlations introduced by transforming the pseudo-data becomes degenerate under the derivative. 
This does not affect our investigation, since we know the central value of our final quasi-PDF will reproduce the original CT14 one. It is expected that the real (imaginary) derivative matrix elements are antisymmetric (symmetric) with respect to $z=0$. The peaks are sharper for larger $P_z$.

The quasi-PDFs using the derivative method (and a filter $F(32,1)$) are shown in the bottom-left of Fig.~\ref{fig:derivative}. The derivative-method PDFs have not only no unphysical oscillation without any filter function needed, but also recover most of the positive-$x$ region parton distribution for $P_z > 4\pi/L$. For the antiquark region, however, we find the sensitive $P_z$ dependence in recovering the small-$x$ PDFs. With the largest nucleon momentum $P_z=24\pi/L$, we can recover the antiquark distribution up to $x=-0.05$. If we apply a soft filter or increase $z_\text{lim}$, it does not make noticeable changes in the distribution with $P_z > 4\pi/L$. 
Lastly, we compare the PDF obtained from the filter and derivative methods, shown in the bottom-right of Fig.~\ref{fig:derivative} using $P_z=24\pi/L$. The derivative method (slightly more complicated to implement) recovers the smaller-$x$ region better: 0.1 vs 0.05. Since real lattice data will have worse signal-to-noise ratios than these pseudo-data, one in principle should try both approaches to see which one works better with the data qualities, reliability of matrix elements obtained in larger $z$ regions, and other issues to address.

\begin{figure*}[htbp]
\includegraphics[width=.4\textwidth]{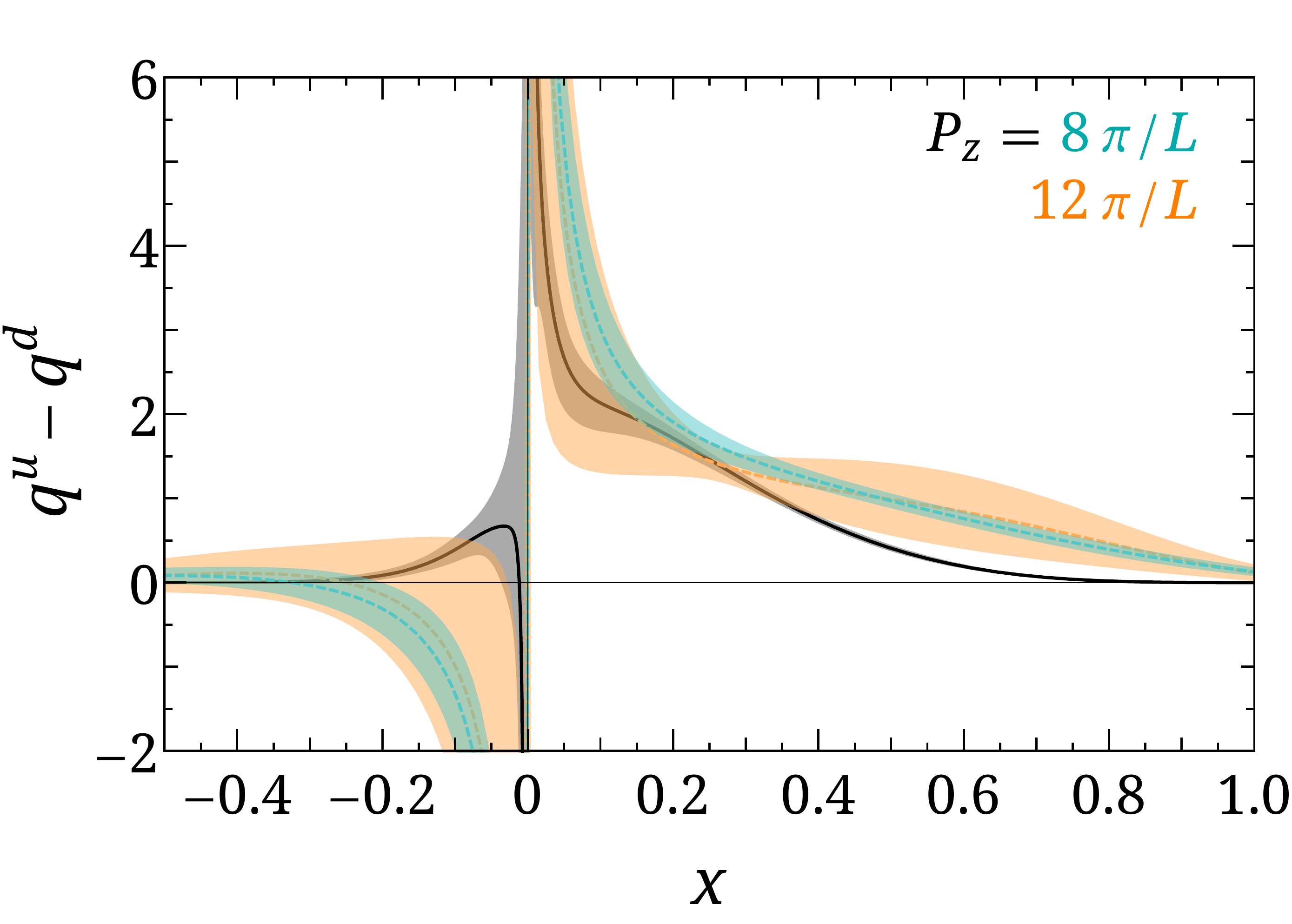}
\includegraphics[width=.4\textwidth]{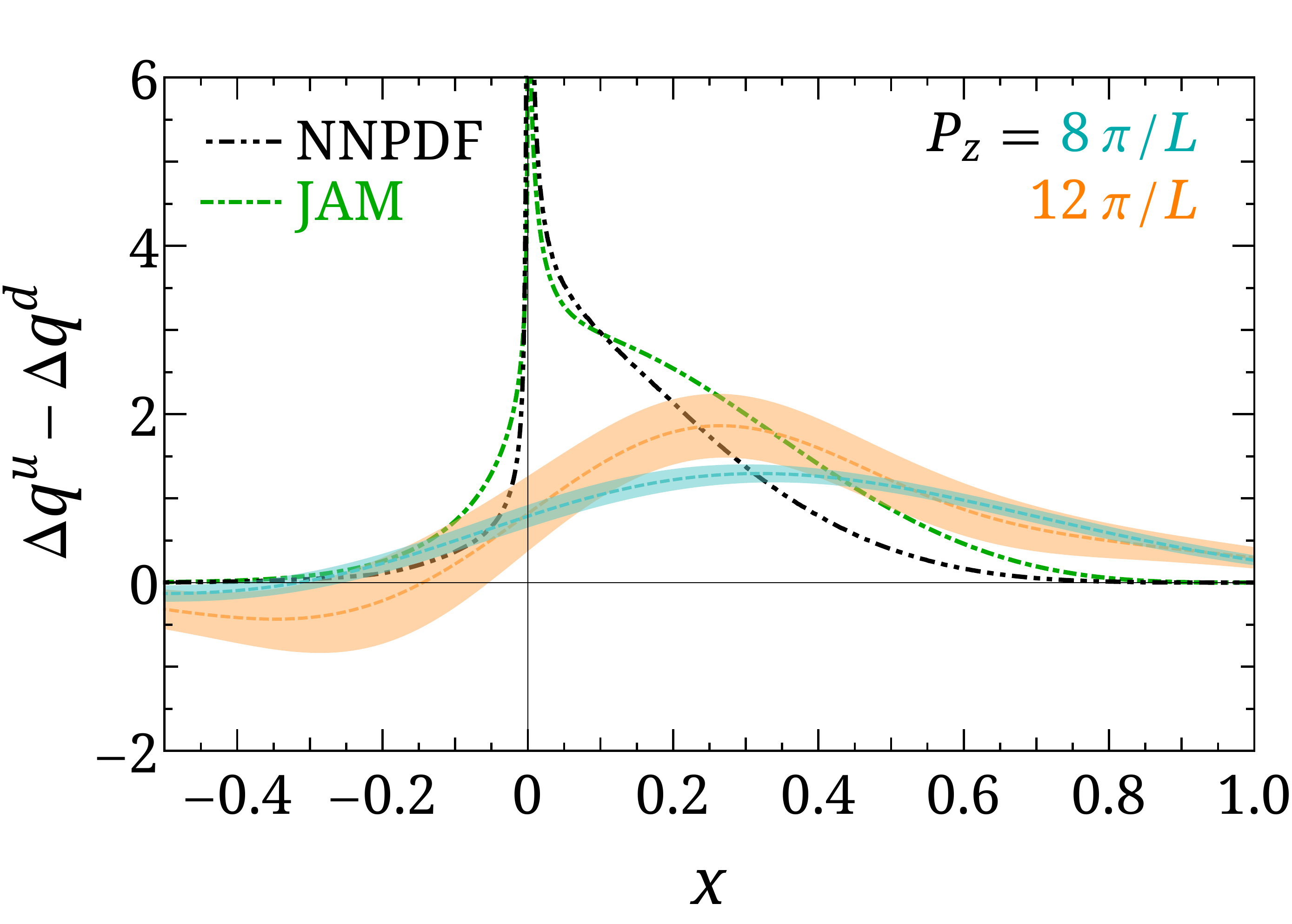}
\caption{The unpolarized (left) and polarized (right) isovector nucleon PDFs as functions of $x$; in both cases, a comparison with global-fit PDFs of the corresponding distribution from largest nucleon boosted momentum $P_z=\{8,12\}\pi/L$. The unpolarized distribution is plotted using the derivative method while the helicity distribution is obtained from the  filter method; both results look promising. Higher momenta are needed to achieve more consistent results.
} 
\label{fig:latticePDF}
\end{figure*}

We apply the improved PDFs methods to the lattice nucleon matrix elements from clover-on-HISQ lattices~\cite{Follana:2006rc,Bazavov:2012xda} and concentrate on the $a=0.09$~fm 135-MeV pion-mass 
ensemble with lattice spatial length 5.76~fm. The nucleon matrix elements are generated using $P_z=\{4,8,12\}\pi/L$ with a $2\times 2$ matrix of standard Gaussian sources and statistics of 2562 measurements each. Two values of source-sink separation, 0.9~fm and 1.08~fm, are used to account for the excited-state contamination using two-state fits~\cite{Lin:2008gv,Bhattacharya:2013ehc}. 
The rest of the unpolarized and polarized lattice calculations and RI/MOM-scheme setup are similar to our previous calculations~\cite{Lin:2014zya,Chen:2016utp,Chen:2017mzz}. 
Figure~\ref{fig:latticePDF} shows results from the quark density (unpolarized) and helicity (polarized) distributions along with selected PDFs, CT14 NNLO~\cite{Dulat:2015mca} at 2~GeV, NNPDFpol1.1~\cite{Nocera:2014gqa} and JAM~\cite{Jimenez-Delgado:2013boa} at 1~GeV. To demonstrate that these two methods work for real lattice data, we use derivative (filter) method for unpolarized (polarized) PDFs. Our previous work~\cite{Chen:2017mzz} shows the RI/MOM matching to $\overline{\text{MS}}$ scheme at 2~GeV is close to 1, and the distribution at 1--2~GeV is small compared with the signal-to-noise ratios here.    
In both cases, the distribution in the larger-$x$ region is slightly larger than the global PDFs; it may not be surprising to see this outcome, since the leading moments $\langle x\rangle$ and $\langle\Delta x\rangle$ from current lattice calculations are around 50\% larger than global-fit PDF values~\cite{Bali:2014gha}. This seems to be what we observe here too. Note that $d(x)/u(x)$ in the global PDF analysis is still an issue in the larger-$x$ region where the uncertainties are larger and there are discrepencies between different groups. It will take effort from both communities to resolve this discrepancy in moment calculations. 
We plan to study $P_z=24\pi/L$ matrix elements in a future calculation and increase the statistics by a couple orders of magnitude to improve upon the current calculations. 

\vspace*{1cm}
\section*{Acknowledgments}
We thank the MILC Collaboration for sharing the lattices used to perform this study. The LQCD calculations were performed using the Chroma software suite~\cite{Edwards:2004sx}. 
This research used resources of the National Energy Research Scientific Computing Center, a DOE Office of Science User Facility supported by the Office of Science of the U.S. Department of Energy under Contract No. DE-AC02-05CH11231;
facilities of the
USQCD Collaboration, which are funded by the Office of Science of the U.S. Department of Energy,
and supported in part by Michigan State University through computational resources provided by the Institute for Cyber-Enabled Research. 
HL thanks Tie-Jiun Hou from CTEQ collaboration for providing the CT14 NNLO isovector PDFs. We thank Luchang Jin, Yi-Bo Yang and Yong Zhao for useful discussions. 
The work of HL is supported by the National Science Foundation under Grant No. 1653405.
JWC is partly supported by the Ministry of Science and Technology, Taiwan, under Grant Nos. 105-2112-M-002-017-MY3, 104-2923-M-002-003-MY3, the CASTS of NTU, the Kenda Foundation, and DFG and NSFC (CRC 110).
TI is supported by Science and Technology Commission of Shanghai Municipality (Grants No. 16DZ2260200) and in part by the Department of Energy, Laboratory Directed Research and Development (LDRD) funding of BNL, under contract DE-EC0012704.
JZ is supported by the SFB/TRR-55 grant ”Hadron Physics from Lattice QCD” and National Science Foundation of China (Grants No. 11405104).

\input{2017PDF-physical.v2.bbl}
\end{document}

%% file: 2017PDF-physical.v2.bbl
%